\def \be {\begin{equation}}
\def \ee {\end{equation}}
\def \bea {\begin{eqnarray}}
\def \eea {\end{eqnarray}}

\def \la {\langle}
\def \ra {\rangle}

\def \del {\partial}
\def \dels {\partial\kern-.5em / \kern.5em}
\def \As {{A\kern-.5em / \kern.5em}}
\def \Ds {D\kern-.7em / \kern.5em}

\def \a {\alpha}
\def \b {\beta}
\def \dag {\dagger}
\def \g {\gamma}

\def \eps {\epsilon}

\def \th {\theta}

\def \star {\ast}

\documentclass[12pt]{article}

\begin{document}
\begin{titlepage}

\begin{center}
\hfill hep-th/0110191\\
\vskip .5in

\textbf{\large
Noncommutative Quantum Mechanics from \\
Noncommutative Quantum Field Theory
}

\vskip .5in
{\large Pei-Ming Ho$^1$, Hsien-Chung Kao$^2$}
\vskip 15pt

{\small \em $^1$Department of Physics,
National Taiwan University, Taipei, Taiwan, R.O.C.}\\
{\small \em $^2$Department of Physics,
Tamkang University, Tamsui, Taiwan, R.O.C.}

\vskip .2in
\sffamily{
pmho@phys.ntu.edu.tw\\
hckao@mail.tku.edu.tw
}

\vspace{60pt}
\end{center}
\begin{abstract}

We derive noncommutative
multi-particle quantum mechanics
from noncommutative quantum field theory
in the nonrelativistic limit.
Paricles of opposite charges are found to
have opposite noncommutativity.
As a result, there is no noncommutative correction
to the hydrogen atom spectrum at the tree level.
We also comment on the obstacles to take
noncommutative phenomenology seriously,
and propose a way to construct noncommutative
$SU(5)$ grand unified theory.

\end{abstract}
\end{titlepage}
\setcounter{footnote}{0}

\section{Introduction}

Recently there has been a growing interest in
noncommutative geometry as well as
its phenomenological implications.
This was motivated by the discovery in string theory
that the low energy effective theory of D-brane
in the background of NS-NS $B$ field 
lives on noncommutative space \cite{CDS}-\cite{SW}.
In the brane world scenerio \cite{AH},
our spacetime may be the worldvolume of a D-brane,
and thus may be noncommutative.
In fact, apart from string theory,
it has long been suggested that the spacetime
may be noncommutative as a quantum effect of gravity,
and it may provide a natural way to regularize
quantum field theories \cite{Snyder,Yang}.

In many proposals to test the hypothetical
spacetime noncommutativity,
one does not need the exact quantum field theory,
but only its quantum mechanical approximation.
Although noncommutative quantum mechanics (NCQM)
has been extensively studied \cite{CST}-\cite{Aca},
we want to clarify a point that has not been emphasized before,
or was even mistunderstood in some of these papers.
The main point is that the noncommutativity
$\th^{ab}$ is not the same for all particles in NCQM.
The noncommutativity of a particle should be opposite to
(differ by a sign from) that of its anti-particle;
and the noncommutativity of a charged particle
should be opposite to any other particle of opposite charge.
Our basic assumption is that
NCQM should be viewed as an approximation of
a noncommutative field theory (NCFT)
in which all fields live on the same noncommutative space.
The same viewpoint was taken in \cite{Bak}.

We will always assume that the time coordinate
$t$ is commutative.
Otherwise the formulation of quantum mechanics
may require drastic modification \cite{Li}.

In Sec. \ref{twoparticle}
we illustrate some ambiguities in defining NCQM.
To resolve these ambiguities we derive NCQM from NCFT
in Sec. \ref{from}.
We find that the noncommutativity of
particle coordinates depends on the charge.
This implies that there is no correction
to the spectrum of the hydrogen atom
due to noncommutativity at the tree level
(Sec. \ref{separation}).
We generalize these results in Sec. \ref{generalization}.
In the last section, we comment on the obstacles
to a complete, consistent description of
noncommutative phenomenology.
Since there exist particles with electric charges
other than $e=0,\pm 1$,
electromagnetic interaction can not be consistently
described as a noncommutative $U(1)$ gauge theory.
Hence we propose a way to construct
noncommutative $SU(5)$ grand unified theory,
where all charges are already properly quantized,
as a better theoretical basis
for noncommutative phenomenology.
In particular, in order to describe
the 10 dimensional antisymmetric representation of $SU(5)$
in the grand unified theory,
we show how to introduce matter fields
in arbitrary representations of the gauge group
using Seiberg-Witten map.

\section{Two-Particle System}\label{twoparticle}

Naively, to define a physical system
on noncommutative space,
we simply take the Lagrangian for ordinary space
and replace all products by star products.
For example, one tends to claim that
the noncommutative Schr\"{o}dinger equation
for a Hydrogen atom is \cite{CST,KD}
\footnote
{
There is an ambiguity in the ordering of the last term.
It could as well be $\psi\ast V$.
However,
replacing $V\ast\psi$ by $\psi\ast V$ is equivalent to
replacing $\th$ by $-\th$.
Without specifying $\th$, we can choose either case
without loss of generality.
}
\be \label{psi}
i\frac{\del}{\del t}\psi=-\frac{\nabla^2}{2m_e}\psi+V(x)\ast\psi,
\ee
where
\be \label{V}
V=-\frac{e^2}{|x|}
\ee
is the electric potential of the proton,
and the $\ast$ product is defined by
\be
f(x)\ast g(x)\equiv e^{\frac{i}{2}\th^{ab}\frac{\del}{\del x^a}
\frac{\del}{\del x^{'b}}}f(x)g(x')|_{x'=x}.
\ee

Here $x$ should be interpreted as the relative coordinate
between the electron and the proton
\be
x^a=x^a_e-x^a_p, \quad a=1,2,3.
\ee
This means that the commutation relation for $x$
should be derived from those for $x_e$ and $x_p$.
Suppose
\be
[x^a_e, x^b_e]=i\th^{ab}_e, \quad
[x^a_p, x^b_p]=i\th^{ab}_p, \quad
[x^a_e, x^b_p]=0,
\ee
then
\be
[x^a, x^b]=i(\th^{ab}_e+\th^{ab}_p).
\ee
We will show below that we should take $\th_e=-\th_p$
and thus $x$ is actually commutative!

If we assume that the proton has infinite mass
and is localized at the origin as a delta function,
we can interpret $x$ as the coordinate of the electron.
Then it would make sense to say that $x$
is a coordinate on the noncommutative space.
However, as it was pointed out in \cite{Bars},
the use of delta function on noncommutative space
invalidates the perturbative expansion in $\th$.
It is also unnatural to assume an extreme localization
of proton on a noncommutative space.

To clarify this problem, we note that
a complete description of the Hydrogen atom should
be given by the total wave function $\Psi(x_e, x_p)$.
On classical space, the Schr\"{o}dinger equation is
\be
i\frac{\del}{\del t}\Psi=\left(
-\frac{\nabla_e^2}{2m_e}-\frac{\nabla_p^2}{2m_p}+V(x_e,x_p)
\right)\Psi,
\ee
where
\be \label{V2}
V(x_e,x_p)=-\frac{e^2}{|x_e-x_p|}.
\ee

When we try to modify this equation to the noncommutative case,
we have to face the following ambiguities.
First, we need to find $V$ on noncommutative space,
and specify the ordering of $V$ and $\Psi$
in the Schr\"{o}dinger equation.
Although this was not a problem for
the wave function of a single particle,
it is a problem for multi-particle states.
The reason is that there is a new possibility
for which the coordinate $x_e$ in $V$
is multiplied from the left,
and $x_p$ in $V$ from the right,
to the wave function $\Psi$.
This can be written as
\be
V\star_{+-}\Psi
\ee
by defining $\star_{\eps_1\eps_2}$ as
\be \label{star+-}
f(x_e,x_p)\star_{\eps_1\eps_2}g(x_e,x_p)\equiv
e^{\frac{i}{2}\th^{ab}\left(
\eps_1\frac{\del}{\del x_e^a}\frac{\del}{\del x_e^{'b}}+
\eps_2\frac{\del}{\del x_p^a}\frac{\del}{\del x_p^{'b}}
\right)}f(x_e,x_p)g(x'_e,x'_p)|_{x=x'}.
\ee

The star product (\ref{star+-}) assumes that
$x_e$ commutes with $x_p$,
although it is mathematically consistent
to assume that $x_e$ does not commute with $x_p$.

To fix these ambiguities,
the basic assumption in our discussion below
is that NCQM is a nonrelativistic approximation of NCFT
in which all fields live on the same noncommutative space.
Without this assumption, the proton and electron
coordinates may have arbitrary independent noncommutativity.

\section{NCQM from NCFT}\label{from}

Consider the NCFT of some charged particles and a $U(1)$ gauge field.
The action is of the form
\be
S=\sum_{\a}S_{\a} + S_A.
\ee
$S_{\a}$ is the action for a charged particle.
For instance, for a fermion
in the fundamental representation of the gauge group,
it is
\be \label{ferm}
S_{\a}=\int d^4 x \bar{\psi}_{\a}\ast(i\Ds+m_{\a})\ast\psi_{\a},
\ee
where $m_{\a}$ is the mass of the particle $\a$ and
\be
D_{\mu}=\del_{\mu}+A_{\mu}.
\ee
The action for the $U(1)$ gauge field is
\be
S_A=\int d^4 x F_{\mu\nu}\ast F^{\mu\nu},
\ee
where
\be
F_{\mu\nu}=[D_{\mu}, D_{\nu}]_{\ast}.
\ee
On a noncommutative space, even the $U(1)$ gauge group
is non-Abelian.
Therefore all fields must have the same charge:
particles have charge $+1$ and anti-particles have charge $-1$.

In order to derive NCQM from NCFT,
we repeat what we do for the commutative case.
First, we collect those terms in the action involving $A_\mu$
\be \label{JA}
\int d^4 x (J^{\mu}\ast A_{\mu}) + S_A,
\ee
where $J=\sum_{\a}j_{\a}$,
and the current density for the fermion in (\ref{ferm}) is
\be
j_{\a}^{\mu}=i\psi_{\a}\ast(\g^0\g^{\mu})^T\ast\psi_{\a}^{\dag}.
\ee
Now we can integrate out $A_\mu$ and find the effective interaction
between the charged particles.
In the weak coupling limit or weak field limit where
we can ignore the self-interaction of $A_\mu$,
one finds the effective interaction
\be \label{SI}
S_I = \int d^4 x d^4 x' J^{\mu}(x)\ast G_{\mu\nu}(x,x')\ast' J^{\nu}(x')
= \int dt H_I,
\ee
where $G$ is the photon propagator in a certain gauge,
and $\ast'$ means $\ast$-product with respect to $x'$.

In fact, we can ignore the $\ast$'s
between $J$ and $A_\mu$ in (\ref{JA}) because
\be
\int d^4 x f(x)\ast g(x)=\int d^4 f(x)g(x).
\ee
We can also drop the $\ast$'s in (\ref{SI}),
and $G$ is simply the usual propagator on commutative space.


Decomposing each field into positive and negative frequency modes
\be
\psi=\int d^3 k(b_{ks}(t)u_{ks}e^{ik_{i}x^{i}}
+d^{\dag}_{ks}(t)v_{ks}e^{-ik_{i}x^{i}}),
\ee
where $b$ is the annihilation operator for the particle,
$d^{\dag}$ is the creation operator for its anti-particle,
and the particle index $\a$ is suppressed.
We will ignore the spinor index $s$
as it will not play a role in our problem.
In the operator formulation,
one can define the field operators
\be \label{field-op+}
\hat{\psi}_+\equiv\int d^3 k b_k(t) e^{ik_{i}x^{i}}
\ee
for the particle $\a$, and
\be \label{field-op-}
\hat{\psi}_-\equiv\int d^3 k d_k(t) e^{ik_{i}x^{i}}
\ee
for its anti-particle $\bar{\a}$.
The quantum mechanical wave function for
a two-particle state $|\xi\ra$ in the NCFT is
\be \label{wavefx}
\Psi_{(\a\eps_1)(\b\eps_2)}(x_1,x_2)\equiv
\la 0|\hat{\psi}_{\a\eps_1}(x_1)\hat{\psi}_{\b\eps_2}(x_2)|\xi\ra.
\ee
Here $x_1$, $x_2$ are viewed as commutative coordinates
in the star product representation.
Thus the coordinates for different particles
in the wavefunction $\Psi$ always commute with one another
by definition.
Similarly, one can define the wave function
for a state of an arbitrary number of particles and anti-particles.

The Schr\"{o}dinger equation is a result of
the fact that $\psi_{\a}$ satisfies its equation of motion,
which can be written as
\be
i\dot{\psi}(x)=[H,\psi(x)]
\ee
in terms of the Hamiltonian $H$.
For the effective action, $H$ is
\be
H=H_0 - H_I,
\ee
where $H_0$ is the kinetic term and $H_I$ is given by (\ref{SI}).
Thus, for example,
\be
i\frac{\del}{\del t}\Psi_{AB}=\la 0|[H,\hat{\psi}_A\hat{\psi}_B]|\xi\ra,
\ee
where $A=(\a\eps_1)$ and $B=(\b\eps_2)$.

Straightforward derivation shows that,
in the non-relativistic approximation
where the interaction is dominated by the Coulomb potential,
the Schr\"{o}dinger equation is given by
\be \label{Schro}
i\frac{\del}{\del t}\Psi_{AB}(x_1,x_2)=\left(
-\frac{\nabla_1^2}{2m_{\a}}-\frac{\nabla_2^2}{2m_{\b}}+V(x_1,x_2)
\right)\star_{\eps_1\eps_2}\Psi_{AB}(x_1,x_2),
\ee
where $\star_{\eps_1\eps_2}$ is defined in (\ref{star+-}),
and $V$ given by (\ref{V2}).

While the above prescription applies to generic interactions,
for our special case of a gauge field,
the result above (\ref{Schro}) can be easily obtained
by demanding gauge symmetry.
For a field in the fundamental representation,
\bea
& \hat{\psi}_+\rightarrow U \ast\hat{\psi}_+, \\
& \hat{\psi}_-\rightarrow \hat{\psi}_-\ast U^{\dag}
\eea
under a gauge transformation.
This implies that
the covariant derivative must act on the wave function $\Psi$
from the left for particles and
from the right for anti-particles.
Since the electric potential $V$ is just
the time component of the gauge potential $A_\mu$,
we immediately reach the same conclusion as in (\ref{Schro}).

Since all fields must have the same charge
in a noncommutative gauge theory,
it is equivalent to say that the coordinate
of each particle of positive (negative) charge
in $V$ is multiplied to the wave function by
the star product with parameter $\th$ ($-\th$).

In the context of string theory,
for an open string ending on a D-brane,
the two endpoints appear as opposite charges
to the D-brane gauge field.
In a $B$ field background,
the two endpoints also observe
opposite noncommutativity \cite{CH}.
It was first argued in \cite{CH} that
the NCFT on a single noncommutative space
automatically takes care of this effect.
In this section we provided a rigorous derivation.

It is interesting to note that although
it is a matter of pure convention whether one uses
$\psi$ to represent, say, the electron or positron,
once we have made the choice,
there is no more freedom in making this choice for
any other fields living on the same noncommutative space.
Here we also see that the charge conjugation
results in a change of noncommutativity
$\th\rightarrow -\th$ \cite{SJ}.

\section{Separation of Variables} \label{separation}

To solve the Schr\"{o}dinger equation for
multi-particle wave functions,
we use the technique of separation of variables.
For the Hydrongen atom,
the Schr\"{o}dinger equation is
\be \label{Sch1}
i\frac{\del}{\del t}\Psi(x_e,x_p)=\left(
-\frac{\nabla_e^2}{2m_e}-\frac{\nabla_p^2}{2m_p}+V(x_e,x_p)
\right)\star_{-+}\Psi(x_e,x_p),
\ee
where we choose the convention that
the noncommutativity parameter $\th$ ($-\th$)
is associated with positive (negative) charges.

Since the kinetic term is not modified,
we take the ansatz
\be
\Psi(x_e,x_p)=\Phi(X)\psi(x),
\ee
where
\be
X=\frac{m_e x_e+m_p x_p}{m_e+m_p}
\ee
is the center of mass (COM) coordinate,
and
\be
x=x_e-x_p
\ee
is the relative coordinate.
The noncommutativity for these coordinates is given by
\bea
\left[ X^i, X^j \right]_{\star_{-+}}&=&
i\frac{m_p-m_e}{m_p+m_e}\th^{ij}
\equiv i\th^{ij}_{pp}, \\
\left[ x^i, x^j \right]_{\star_{-+}}&=&0, \\
\left[ x^i, X^j \right]_{\star_{-+}}&=&
i\th^{ij}
\equiv i\th^{ij}_{ep}.
\eea
The kinetic term can be rewritten as
\be
\frac{\nabla_e^2}{2m_e}+\frac{\nabla_p^2}{2m_p}=
\frac{\nabla_X^2}{2M}+\frac{\nabla_x^2}{2m},
\ee
where
\be
M=m_e+m_p
\ee
is the total mass and
\be
m=\frac{m_e m_p}{m_e+m_p}
\ee
is the reduced mass.

For the Fourier mode of $X$,
\be
\Psi(X)=e^{-iE t+iK_i X^i}\psi(x),
\ee
(\ref{Sch1}) is reduced to
\be \label{Sch2}
\left(E-\frac{K^2}{2M}\right)\psi(x)=\left(-\frac{\nabla_x^2}{2m}+
V(x-\frac{1}{2}\th_{ep}K)\right)\psi(x).
\ee
Note that translational invariance implies that
$V$ can only depend on the relative coordinate $x$.
Let $\psi(x)=\psi'(x-\frac{1}{2}\th_{ep}K)$.
Since (\ref{Sch2}) contains no star product,
it is exactly the same equation for classical space
in terms of $\psi'$.
Unless we include self-interactions of the gauge field,
the whole spectrum is exactly the same as the commutative case!
The shift in the relative coordinate is easy to understand from the D-brane picture,
where space non-commutitivity is resulted from background $B$ field.

Therefore, for example,
the noncommutative correction to Lamb shift
should be much smaller than the one given in \cite{CST}.
There is no correction at tree level.
The lowest order contribution of $\th$ comes from
the one-loop diagrams and is negligible.

\section{Generalization} \label{generalization}

In \cite{CPST} it was shown that a matter field
in the fundamental representation is not allowed
to couple to more than two different gauge fields
on noncommutative space.
So far we have only considered the case of one gauge field.
It is straightforward to include another gauge field.

Suppose that there are $m$ particles.
Let the charges of particle $\a$ ($\a=1,\cdots,m$) be
$q^{\a}$, where $q^{\a}=1, 0, -1$.
If $q^{\a}=1$, it means that the field operator
for particle $\a$,
which was denoted as $\hat{\psi}_+$ before,
transforms from the left
\be
\phi_{\a}\rightarrow U\ast\phi_{\a}.
\ee
It transforms from the right
\be
\phi_{\a}\rightarrow \phi_{\a}\ast U
\ee
if $q^{\a}=-1$.
The covariant derivative of a field operator is
\be
D_{\mu}\phi_{\a}=\del_{\mu}\phi_{\a}
+q^{\a} A_{\mu}\star_{q^{\a}}\phi_{\a},
\ee
where $\star_{\pm}$ is the star products
with the parameter $\pm\th$.

If we repeat the derivation in the previous sections,
the Schr\"{o}dinger equation for $N$ particles is
\be
i\frac{\del}{\del t}\Psi(x_1,\cdots,x_N)=
-\sum_{\a=1}^{N}\frac{\nabla_i^2}{2m_i}\Psi
+\frac{1}{2}\sum_{\a\neq\b}q_{\a}q_{\b}V(x_{\a}, x_{\b})
\star_{q_{\a}q_{\b}}\Psi,
\ee
where $V$ is the $(00)$ component of the Green's function
for the gauge field $A_\mu$.

The COM coordinates $X^{\mu}$ of the system satisfies
\be
[X^{\mu}, X^{\nu}]_{\star_{ \{q_i\} }}=
i\frac{\sum_{\a=1}^{N}q_i^{\a}m_{\a}^2}{\sum_{\b}m_{\b}^2}\th^{\mu\nu}.
\ee
It is easy to see that the magnitude of the noncommutativity
is never larger than $|\th|$.

A composite particle is a system of $N$ particles
which has a bound state with a small spatial extension.
The COM coordinates of the system will be taken
as the coordinates of the composite particle.
If the size of the composite particle
is larger than $\sqrt{\th}$,
it is meaningless to talk about its noncommutativity.
In the case of a Hydrogen atom,
the relative coordinates $x$ is commutative,
thus its size can be arbitrarily small.
On the other hand, if some relative coordinates
for the constituents of the composite particle
are noncommutative,
which is always the case as long as there are
three or more charged constituent particles,
the size of the composite particle must be larger than
the order of $\sqrt{\th}$,
and hence the noncommtativity of the composite particle
can be neglected for most purposes.

\section{Discussion}

On noncommutative (NC) space,
charges are always quantized,
even for $U(1)$ gauge field.
However, in the standard model,
there are particles of electric charges $1/3, 2/3$ etc.
It implies that the electromagnetic interaction
can not be a NC $U(1)$ gauge theory.
Similarly, the $U(1)$ gauge group for hypercharges
can not be noncommutative, either \cite{CPST2}.
In the $SU(5)$ Grand Unified Theory (GUT),
on the other hand,
all charges are already quantized.
There are fractional hypercharges only because
the $U(1)$ group is embedded in $SU(5)$ with
a generator $T=\mbox{diag}(1/3,1/3,1/3,-1/2.-1/2)$.
But there are other problems for NCGUT.
The first problem is to define NC $SU(5)$ gauge symmetry.
In general, it is straightfoward to construct
NC $U(N)$ gauge theory,
but difficult to have any other gauge group
\cite{Terashima,Armoni,CPST}.

A possible resolution of this problem \cite{JMSSW}
is to define NC $SU(N)$ gauge symmetry as the image of
the classical $SU(N)$ via Seiberg-Witten (SW) map \cite{SW}
\be
\hat{A}=\hat{A}(A),
\ee
where quantities without (with) hats are
commutative (noncommutative) fields.
It is consistent with gauge transformations
to restrict $A_\mu$ to the Lie algebra of $SU(N)$.
The same idea can be used to define the noncommutative
version of any classical group \cite{JMSSW}.


It is also possible to define NC $SU(5)$ theory
directly in terms of the noncommutative variable $\hat{A}$
without mentioning the commutative $A$.
We can simply take the NC $U(N)$ gauge field $\hat{A}$
and impose the following constraint
\be
C_{\mu\nu}(k)\equiv \mbox{Tr}F_{\mu\nu}(\hat{A})(k)=0,
\ee
where $F_{\mu\nu}(\hat{A})$ is the inverse SW map.
(An exact expression for the inverse SW map
was given in \cite{Liu2,OO}.)
This implies that the $U(1)$ part of $A_\mu$ can be gauged away,
and the result is equivalent to the approach of \cite{JMSSW}.

It is interesting to note that another constraint
with a much simpler expression
\be
C'_{\mu\nu}(k)\equiv
\mbox{Tr}\int d^4 x \hat{F}_{\mu\nu}(x)\ast
e_{\ast}^{ik_{\mu}(x^{\mu}+i\th^{\mu\nu}\hat{A}_{\nu})}=0
\ee
is also gauge invariant and has the same classical limit
$\mbox{Tr}F_{\mu\nu}=0$.
At this moment we do not know if these two constraints
are exactly the same.


Recently, a similar idea was proposed independently in \cite{CD},
where the constraint was imposed on $\hat{A}$ instead.
Another constraint on gauge transformations $\hat{U}$
has to be imposed simultaneously for consistency \cite{CD}.
It would be of interest to know if all such constraints
are equivalent under field redefinitions.

Another problem about NCGUT is that
there are matter fields in
the antisymmetric representation of $SU(5)$.
Althgouh it is straightforward to
define matter fields in the fundamental
and adjoint representations,
in general it is hard to introduce other representations
\cite{Terashima,CPST}.

This problem can also be solved by using the SW map.
For any $D$ dimensional representation of $SU(5)$,
we consider the SW map for NC $U(D)$ gauge symmetry.
For any classical gauge transformation $U\in U(D)$,
the SW map provides a NC $U(D)$ transformation $\hat{U}(U)$.
Since $SU(5)$ can be embedded in $U(D)$
according to its $D$ dimensional representation,
we can define NC $SU(5)$ transformations for
a fundamental representation of $U(D)$ by
\be
\hat{\phi}_a \rightarrow \hat{U}_{ab}(U)\hat{\phi}_b,
\quad a,b = 1, 2, \cdots, D,
\ee
where $U$ is a classical $SU(5)$ gauge transformation.
Thus $\hat{\phi}$ can be viewed as a $D$ dimensional
representation of NC $SU(5)$.

For the 10 dimensional representation of $SU(5)$,
one usually defines it as an antisymmetric tensor
$\phi_{ij}=-\phi_{ji}$ $(i,j=1, 2, \cdots, 5)$ which transforms as
$\phi\rightarrow U\phi U^{\dag}$.
However, the tensor will not be antisymmetric
after a generic NC gauge transformation.
In the above we avoided this problem by
defining this representation directly
as a 10 component colume without any constraint.

Similarly, it is consistent with classical $SU(5)$ gauge transformations
to restrict the classical $U(D)$ gauge potential $A^{(D)}$
to the $su(5)$ Lie algebra  embedded in $u(D)$.
Its image under the SW map can be viewed as the NC gauge potential
in the $D$ dimensional representation of NC $SU(5)$.
The covariant derivative of a matter field in
this representation is
\be
\hat{D}_{\mu}\hat{\phi}=(\del_{\mu}+\hat{A}^{(D)}_{\mu}(A))\hat{\phi},
\ee
where $A$ is the commutative $SU(5)$ gauge potential.
Obviously, this construction also works for other gauge groups.

Finally, due to the UV-IR mixing,
the UV divergences of NC quantum field theories result in
new IR poles nonperturbative in $\th$ \cite{MRS,RS}.
For a comprehensive discussion on this problem see \cite{DN}.
In order to give a reliable, consistent description of
NC electromagnetic interactions,
or any other low energy phenomena on NC space,
it is necessary to properly address all these problems.
We leave these issues for future study.

\section*{Acknowledgment}

The authors thank Chong-Sun Chu, Xiao-Gang He
and Staphen Narison for helpful discussions.
This work is supported in part by
the National Science Council,
the Center for Theoretical Physics
at National Taiwan University,
the National Center for Theoretical Sciences,
and the CosPA project of the Ministry of Education,
Taiwan, R.O.C.

\end{document}